\documentclass[pra,twocolumn,aps,floatfix,showpacs,tightenlines,superscriptaddress,amsmath,amssymb,showkeys,10pt,nofootinbib]{revtex4-1}
\usepackage{amssymb,amsmath,amsthm,amsbsy,epsfig,color,graphicx}
\usepackage[ansinew]{inputenc}
 \usepackage[english]{babel}

\usepackage{mathtools}
\usepackage{tikz}

\usepackage{color}
\definecolor{myblue}{rgb}{.8, .8, 1}

\usepackage{amsmath, amssymb, bbold}
\usepackage{amsfonts}
\usepackage{empheq}
\usepackage[pdftex,
            pdfauthor={Kyrylo Simonov},
            pdftitle={Particle decay and the emergence of classicality: Spontaneous collapse case}]{hyperref}
\hypersetup{colorlinks=true,urlcolor=red,citecolor=blue,filecolor=magenta,bookmarks=true}

\newlength\mytemplen
\newsavebox\mytempbox

\makeatletter
\newcommand\mybluebox{%
    \@ifnextchar[
       {\@mybluebox}%
       {\@mybluebox[0pt]}}

\def\@mybluebox[#1]{%
    \@ifnextchar[
       {\@@mybluebox[#1]}%
       {\@@mybluebox[#1][0pt]}}

\def\@@mybluebox[#1][#2]#3{
    \sbox\mytempbox{#3}%
    \mytemplen\ht\mytempbox
    \advance\mytemplen #1\relax
    \ht\mytempbox\mytemplen
    \mytemplen\dp\mytempbox
    \advance\mytemplen #2\relax
    \dp\mytempbox\mytemplen
    \colorbox{myblue}{\hspace{1em}\usebox{\mytempbox}\hspace{1em}}}

\makeatother

\usepackage[framemethod=tikz]{mdframed}
\definecolor{mycolor}{rgb}{0.122, 0.435, 0.698}
\newmdenv[innerlinewidth=0.5pt, roundcorner=4pt,linecolor=mycolor,innerleftmargin=6pt,
innerrightmargin=6pt,innertopmargin=6pt,innerbottommargin=6pt]{mybox}
\usepackage{paracol}
\usepackage{tcolorbox}
\tcbuselibrary{theorems}

\newtcbtheorem{Definitions}{Assumption}%
{colback=mycolor!5,colframe=mycolor,fonttitle=\bfseries,
left=4pt,
right=4pt,
top=5pt,
bottom=5pt}{defi}

\newtcbtheorem{Lemmas}{Lemma}%
{colback=green!5,colframe=green!50!blue,fonttitle=\bfseries,
left=4pt,
right=4pt,
top=5pt,
bottom=5pt
}{lemma}

\newtcolorbox[blend into=figures]{boxdefi}[3][]
{ float*=ht,width=\textwidth,lower separated=false, center upper,
title={#2},label= def:#3,#1}

\definecolor{darkblue}{rgb}{0.0, 0.33, 0.71}

\begin{document}

\title{Particle mixing and the emergence of classicality: A spontaneous collapse model view}

\author{Kyrylo Simonov}
\email{kyrylo.simonov@univie.ac.at}
\affiliation{Fakult\"{a}t f\"{u}r Mathematik, Universit\"{a}t Wien, Oskar-Morgenstern-Platz 1, 1090 Vienna, Austria}

\begin{abstract}
Spontaneous collapse models aim to resolve the measurement problem in quantum mechanics by considering wave-function collapse as a physical process. We analyze how these models affect a decaying flavor oscillating system whose evolution is governed by a phenomenological non-Hermitian Hamiltonian. In turn, we apply two popular collapse models, the QMUPL and the CSL models, to a neutral meson system. By using the equivalence between the approaches to the time evolution of decaying systems with a non-Hermitian Hamiltonian and a dissipator of the Lindblad form in an enlarged Hilbert space, we show that spontaneous collapse can induce the decay dynamics in both quantum state and master equations. Moreover, we show that the decay property of a flavor oscillating system is intimately connected to the time (a)symmetry of the noise field underlying the collapse mechanism. This (a)symmetry, in turn, is related to the definition of the stochastic integral and can provide a physical intuition behind the It\=o/Stratonovich dilemma in stochastic calculus.
\end{abstract}



\date{\today}

\maketitle

\section{Introduction\label{intro}}

Despite a distinctly high success of (``standard'') quantum mechanics in describing the microscopic world and covering plenty of phenomena on different energy scales, it meets some conceptual controversies. The linearity of the Schr\"{o}dinger equation manifests itself at the famous superposition principle, which is one of the cornerstones of quantum mechanics. At high energy scales, it plays a crucial role in the phenomena of particle mixing and oscillations, which are experimentally observed in several systems such as neutral mesons~\cite{Christenson1964, Abashian2001}. They occur when the energy eigenstates of the particle are not necessarily identical to the interaction eigenstates but rather are their superpositions. For example, a neutral K-meson produced in strong interactions as a bound state of a down quark and strange antiquark $d \bar{s}$ can be found to turn into the bound state $\bar{d} s$ via weak interaction processes. However, the superposition principle does not seem to be relevant in the macroscopic world. We do not observe a table being here and there and a kitten dead and alive at once: our everyday experience shows up a break of the dynamics predicted by the linear Schr\"{o}dinger equation. This observation leads to a question: how the everyday classical world arises from the quantum world?

Going further, a measurement performed on a quantum system and associated with an interaction between the system and the measurement apparatus (implied to be macroscopic) reveals a definite outcome. In turn, it is intimately connected to the so-called measurement problem, which, in particular, refers to the questions of what selects a certain outcome in a particular experimental run and what makes an actual interaction a measurement (``What makes a measurement a measurement?'')~\cite{Brukner2015, Baumann2018}. The measurement problem and, in general, the validity of the superposition principle is a subject of intense experimental verification.

``Standard'' quantum mechanics does not explain this quantum-to-classical transition but only postulates an ad hoc separation between microscopic (quantum) and macroscopic (classical) worlds. This results in two different types of dynamics of the quantum system,
\begin{itemize}
 \item a stochastic and non-unitary reduction (``collapse'') of the state of the quantum system due to an interaction with a macroscopic system (``measurement apparatus'') in accordance with the Born's rule,
 \item a deterministic unitary time evolution governed by the Schr\"{o}dinger equation before and after the measurement.
\end{itemize}
A possible approach to the quantum-to-classical transition can be a universal dynamics valid on all scales, which contains both quantum and classical mechanics as approximations. Dynamical reduction models, or so-called collapse models, aim to provide a phenomenological framework to such dynamics. They introduce an ontologically objective mechanism of the wavefunction collapse, which is implemented by replacing the Schr\"{o}dinger dynamics with its stochastic (in order to explain why the measurement outcomes occur randomly in accordance with the Born's rule) and nonlinear (in order to break up a macroscopic superposition and get rid of the Schr\"{o}dinger's cat) modification~\cite{Feldmann2011, Bassi2003, Bassi2012}.

The first and the simplest collapse model is the GRW (Ghirardi--Rimini--Weber) model, also known as the QMSL (Quantum Mechanics with Spontaneous Localizations) model, introduced in 1985~\cite{GRW}. This model has implemented the collapse mechanism by assuming, with respect to a system of $N$ distinguishable particles, random spontaneous localizations (called ``hittings'' by Ghirardi~\cite{Ghirardi1999} and ``jumps'' by Bell~\cite{Bell}) with a mean rate $\lambda$ affecting each particle and conserving the usual Schr\"{o}dinger evolution of the system between the successive localizations. In turn, the GRW model gives rise to the amplification mechanism tuned by the rate $\lambda$, which sets the collapse strength: spontaneous localizations affect a microscopic object very rarely and can be neglected, while a macroscopic superposition is rapidly reduced~\cite{Bassi2003, Bassi2012}. Another guiding line implemented by the GRW model is a choice of the preferred basis, into which the state of the quantum system is reduced so that a macroscopic object has a definite position in space: for that purpose, the GRW model chooses the position basis. This leads to the definition of the collapse width $r_C$, which sets a coarse-graining for the wave function of the system. Basically, it sets a scale at which a spatial superposition is effectively reduced and, therefore, is a threshold between macroscopic and microscopic superpositions~\cite{BassiUlbricht}. For example, a superposition of two localized states separated by a distance $d \ll r_C$ is not significantly affected by the collapse dynamics, whereas one with $d \gg r_C$ will be effectively localized. Taking collapse models seriously, the parameters $\lambda$ and $r_C$ are two new natural constants introduced by these models.

An important class of dynamical reduction models is represented by those that describe the wave-function collapse as a continuous process induced by an interaction between the system and a (classical) noise field. The benefit of these models is a possibility to govern the universal dynamics by a stochastic differential equation, which is obtained, generally speaking, by adding new non-linear and stochastic terms to the Schr\"{o}dinger equation. The typical and, perhaps, most popular models of this class are the QMUPL (Quantum Mechanics with Universal Position Localization)~\cite{QMUPL} and the CSL (Continuous Spontaneous Localization)~\cite{PearleCSL, GhirardiCSL} models. One of the essential differences between these models is the simpler mathematical structure of the QMUPL model: the QMUPL noise field ``lives'' only in the time dimension, whereas the CSL noise field is spread both in time and space. Therefore, the CSL model introduces two parameters, a collapse rate and a coherence length (just like the GRW model), while, in the QMUPL model, the latter is absent. For many physical systems, both models predict approximately equivalent dynamics. In this paper, however, we will provide an example of a setup featuring flavor mixing, for which the QMUPL and the CSL models offer non-equivalent predictions.

Since their appearance on the market collapse models caught the eye of researchers and were intensively investigated in a plethora of physical systems at different energy scales~\cite{CarlessoDonadi}. In particular, collapse models were analyzed with respect to the spontaneous radiation emission from charged particles~\cite{BassiDuerr, BassiDonadi, Adler2013} and put to experimental tests by X-rays~\cite{Curceanu2015, Piscicchia2015, Curceanu2016, Piscicchia2017, Curceanu2019}. Furthermore, spontaneous collapse models were recently studied in the context of cold-atom experiments~\cite{Bilardello2016}, gravitational waves~\cite{Carlesso2016}, levitated nanoparticles~\cite{Vinante2019}, matter-wave interferometry~\cite{Nimmrichter2011, Toros2017, Toros2018, Schrinski2020}, and optomechanical setups~\cite{Carlesso2018, Nobakht2018, Carlesso2019}. 

Particular attention is attracted to the analysis of the possible effects of a spontaneous collapse in the oscillations at high energies. Mixed systems propose some peculiar features, which have stimulated high interest in these systems as a rich playground for testing the very foundations of quantum mechanics~\cite{Genovese2004, Bramon2006, Bernabeu2006, Hiesmayr2008, Yabsley2008, Blasone2008, AmelinoCamelia2010, DiDomenico2012, Hiesmayr2012, Bernabeu2012, Capolupo2019, Durt2019, Simonov2019, Bernabeu2019, Naikoo2020, Buoninfante2020}. For example, the beauty of the neutral K-meson systems, whose relevant states measured in experiments are superpositions of states with distinct masses, is that they reveal a parallel with the spin-$\frac{1}{2}$ particles and photons. However, they offer richer properties such as decay and the violation of the $\mathcal{CP}$ discrete symmetry, which leads to surprising results such as a contradiction between local realism and the $\mathcal{CP}$ violation~\cite{Bertlmann2001}. In this context, a possible effect of spontaneous collapse on flavor oscillations was studied for neutrinos~\cite{Bahrami2013, Donadi2013} and neutral mesons, in particular, $K^0-\bar{K}^0$ correlated pairs produced at the CPLEAR experiment by the LEAR accelerator facility at CERN and at the KLOE and KLOE-2 experiments by the DA$\mathrm{\Phi}$NE
$\phi$-factory at the Frascati National Laboratories of INFN~\cite{Bahrami2013, Donadi2012}. In turn, the CSL model has predicted exponential damping of the oscillatory behavior of a neutral meson system~\cite{Bahrami2013, Donadi2012}.

It has been put under discussion whether the spontaneous collapse can be considered also as a source of the particle decay~\cite{SimonovLetter, SimonovPaper, SimonovKLOE}. It has been shown that the decay dynamics can be recovered by the mass-proportional CSL model with an asymmetric correlation function of the noise field underlying this collapse model. Mathematically, the asymmetry in the noise field is expressed as a parameter in the quantum state equation of the collapse model, which is intimately related to the choice of the stochastic formalism. Considering this asymmetry free can lead to a non-trivial dependence on the absolute masses of the eigenstates of the time evolution. In~\cite{Tilloy2017}, this point of view was criticized by considering a unique fixed master equation in the Gorini--Kossakowski--Lindblad--Sudarshan (GKLS) form with a Hermitian Hamiltonian, which, clearly, does not describe the decay property of particles. This argument combined with the fixed asymmetry of the noise field due to the chosen stochastic formalism led to the conclusion that spontaneous collapse models are not able to induce the (exponential) decay dynamics. This is due to the fact that a na\"{i}ve switch of the stochastic formalism changes both the asymmetry of the noise field and the quantum state equation in such a way that the master equation preserves its form (as it should be), and the non-Hermitian part of the Hamiltonian, which covers the decay, clearly, cannot be recovered in such way.

In this paper, we review the results of~\cite{SimonovLetter, SimonovPaper, SimonovKLOE} in the context of the master equation associated with a collapse model and investigate how the noise field affects the dynamics of neutral mesons. We derive a class of collapse models with a free choice of time-asymmetry of the underlying white noise field and show that the decay property of a flavor oscillating system (or its absence) can be governed by the corresponding spontaneous collapse dynamics in accordance with~\cite{SimonovPaper}. In turn, the derived collapse models generate a family of master equations parameterized by the decay rates, including the GKLS equation corresponding to the evolution without decay.

The paper is organized as follows. In Section~\ref{particles}, we discuss the phenomenology of neutral mesons with putting the focus on the non-relativistic framework via Wigner--Weisskopf approximation and the corresponding master equation. This approach operates with a two-state non-Hermitian Hamiltonian acting on the flavor Hilbert space and results in exponential decay dynamics in the transition probabilities. Furthermore, we discuss the approach to the neutral mesons phenomenology based on the enlarged Hilbert space with included decay product states, which allows to define a Hermitian Hamiltonian and, in turn, prove the completely positive dynamics of a neutral meson system. In Section~\ref{collapsedyn}, we discuss the collapse equation in the flavor Hilbert space for a non-Hermitian Hamiltonian and associate it with the master equation introduced in the previous Section. After all, we generalize this collapse model to a class of models with a time-asymmetric noise field, where the collapse process turns out to be the source of the decay widths. In Section~\ref{qmuplcsl}, we refer to the QMUPL and the mass-proportional CSL models, generalize them to the time-asymmetric noise field introduced in the previous Section, and derive the transition probabilities using the corresponding master equations. 

\section{Phenomenology of the neutral mesons \label{particles}}
Mixed systems such as the $M^0$--$\bar{M}^0$ systems can be described by a phenomenological Hamiltonian acting on a two-dimensional Hilbert space $\mathbf{H}_M$ called also flavor space. The physical (flavor) states $|M^0\rangle$ and $|\bar{M}^0\rangle$ of a neutral meson are labeled by a flavor quantum number and can decay into the same final states. In particular, the flavor states of neutral K-mesons are labeled by the strangeness quantum number ($S=\pm 1$, respectively) and, for the hadronic decays, both can decay via weak interaction into two or three pions. The usual approach to the dynamics of a neutral meson system is based on the Wigner--Weisskopf approximation, which considers an effective non-Hermitian Hamiltonian
\begin{equation}\label{nonHermHam}
    \hat{H} = \hat{M} - \frac{i}{2}\hat{\Gamma},
\end{equation}
with the eigenstates $|M_i\rangle$ (in particular, for neutral mesons, $i = L, H$ corresponding to "light" and "heavy") with the (distinct) definite masses $m_i$ and known decay widths $\Gamma_i$, so that the corresponding eigenvalues read $m_i - \frac{i}{2}\Gamma_i$. In this treatment, $\hat{M} = \hat{M}^\dagger$ is the mass operator, which covers the unitary part of the dynamics, and $\hat{\Gamma} = \hat{\Gamma}^\dagger$ describes the decay. Apart from neutral K-mesons, the difference of the decay widths $\Delta\Gamma = \Gamma_L - \Gamma_H$ in the $M^0$--$\bar{M}^0$ systems is tiny. Up to a particular phase convention and neglecting a slight violation of the $\mathcal{CP}$ symmetry in the weak interaction, the mass (lifetime) eigenstates $|M_i\rangle$ of the Hamiltonian~(\ref{nonHermHam}) can be related to the flavor states as
\begin{equation}
|M_{H/L}\rangle = \frac{1}{\sqrt{2}} (|M^0\rangle \pm |\bar{M}^0\rangle).
\end{equation}
The Wigner--Weisskopf approximation takes into account only the time evolution of the components of the flavor states, so that the neutral meson system dynamics is described by the effective Schr\"{o}dinger equation,
\begin{eqnarray}\label{nonHermSchr}
 i\frac{d}{dt}|\psi\rangle_t &=& \Bigl(\hat{M} - \frac{i}{2}\hat{\Gamma} \Bigr)|\psi\rangle_t, \\
 |\psi\rangle_t &=& a(t) |M^0\rangle + b(t) |\bar{M}^0\rangle,
\end{eqnarray}
where $\hbar = c = 1$ is assumed. Due to the presence of the non-Hermitian part of the Hamiltonian~(\ref{nonHermHam}), the temporal part of the evolution of the neutral meson system is not normalized. Namely, considering an arbitrary state $|\psi\rangle_t \in \mathbf{H}_M$, we obtain
\begin{equation}\label{unnorm}
d |||\psi\rangle_t||^2 = -\sum_i \Gamma_i | \langle M_i | \psi \rangle_t |^2 dt,
\end{equation}
so that $|\psi\rangle_t$ is not normalized when $\Gamma_i \neq 0$. In particular, let us consider the quantities actively investigated at the accelerator facilities, the transition probabilities
\begin{equation}
    P_{\mathrm{in} \rightarrow \mathrm{out}}(t) = |\langle\psi_{\mathrm{out}}|e^{-i\hat{H}t}|\psi_{\mathrm{in}}\rangle|^2
\end{equation}
for the states $|\psi_{\mathrm{in},\mathrm{out}}\rangle \in \mathbf{H}_M$. The most interesting are the transition probabilities for the lifetime eigenstates $|M_i\rangle$,
\begin{equation}\label{ProbMS}
P_{M_i \rightarrow M_j}(t) = e^{-\Gamma_i t} \delta_{ij},
\end{equation}
and the flavor states $|M^0/\bar{M}^0\rangle$,   
\begin{equation}\label{ProbFS}
P_{M^0 \rightarrow M^0/\bar{M}^0}(t) = \frac{1}{4}\Bigl\{ \sum_i e^{-\Gamma_i t} \pm 2 e^{-\Gamma t} \cos[t \Delta m] \Bigr\},
\end{equation}
where $\Gamma = \frac{\Gamma_L + \Gamma_H}{2}$. Since the temporal part of the evolution of the $M^0$--$\bar{M}^0$ system is not normalized due to~(\ref{unnorm}), the probabilities~(\ref{ProbMS}) and~(\ref{ProbFS}) are not conserved. A formal renormalization by the decay width would allow to apply the Born's rule within a framework of time operator, however, this option is falsified in neutral K-meson systems because of the $\mathcal{CP}$ violation~\cite{Durt2019}.

If the $M^0$--$\bar{M}^0$ system is considered as open (i.e., interacting with the external environment) and affected by some uncontrolled phenomena, the resulting evolution can be described by the following master equation~\cite{Bertlmann2006},
\begin{widetext} 
\begin{eqnarray}
\nonumber \frac{d\hat{\rho}_t}{dt} &=& -i\hat{H}\hat{\rho}_t + i\hat{\rho}_t \hat{H}^\dagger - \frac{1}{2} \sum_i [ \hat{L}_i^\dagger\hat{L}_i \hat{\rho}_t + \hat{\rho}_t \hat{L}_i^\dagger\hat{L}_i - 2 \hat{L}_i\hat{\rho}_t \hat{L}_i^\dagger] \\
 &=& i[\hat{\rho}_t, \hat{M}] - \frac{1}{2} \sum_i [ \hat{L}_i^\dagger\hat{L}_i \hat{\rho}_t + \hat{\rho}_t \hat{L}_i^\dagger\hat{L}_i - 2 \hat{L}_i\hat{\rho}_t\hat{L}_i^\dagger] - \frac{1}{2}\{\hat{\Gamma}, \hat{\rho}_t \} \label{GenME},
\end{eqnarray}
\end{widetext}
where $\hat{\rho}_t \in \mathcal{D}(\mathbf{H}_M)$ is the density operator\footnote{The set $\mathcal{D}(\mathbf{H}_M)$ of density operators on $\mathbf{H}_M$ is a convex subset of the space $\mathcal{B}(\mathbf{H}_M)$ of bounded linear operators on $\mathbf{H}_M$ formed by the positive self-adjoint trace-class operators.} which represents the state of the neutral meson system, $\hat{L}_i$ are the Lindblad generators which describe the interaction between the system and environment, and curly brackets denote an anticommutator. In the absence of decay, i.e., when the decay operator $\hat{\Gamma}$ is set to be zero, Eq.~(\ref{GenME}) possesses the GKLS form~\cite{GKS, Lindblad} and, hence, describes a completely positive evolution of the system. This means that the operator $\hat{\rho}_t$ remains a density operator and still represents a state of the quantum system, so that the evolution given by the master equation~(\ref{GenME}) is physically consistent. In particular, it can govern dissipative and decoherence effects in flavor mixing, which can be signals for new physics. For example, decoherence in neutrino oscillations can reveal the difference between Dirac and Majorana neutrinos and, moreover, break up the $\mathcal{CPT}$ symmetry~\cite{Capolupo2019, Buoninfante2020}.

Although the master equation (\ref{GenME}) does not possess the GKLS form if $\hat{\Gamma} \neq 0$, it is possible to prove the completely positive dynamics with respect to the non-Hermitian Hamiltonian~(\ref{nonHermHam}) in an elegant way by considering particle decay as an open system~\cite{Bertlmann2006, Hiesmayr2017}. Namely, the issue of non-conserving probabilities due to~(\ref{unnorm}) can be resolved by taking into account the decay products. This is done by enlarging the Hilbert space $\mathbf{H}_M$ (spanned by the eigenstates of the Hamiltonian~(\ref{nonHermHam})) to the Hilbert space $\mathbf{H} = \mathbf{H}_M \oplus \mathbf{H}_D$, where $\mathbf{H}_D$ is spanned by the orthonormal states $|f_i\rangle$ which represent the decay products~\cite{Bertlmann2006}. In this new space $\mathbf{H}$, the decay property of neutral mesons can be incorporated to a GKLS equation as a Lindblad operator. Physically, this means that the decay property is induced by an interaction of the neutral meson system with an environment (analogously to an interaction with the QCD vacuum in QFT), and the resulting time evolution is completely positive~\cite{Bertlmann2006, Hiesmayr2010}. We construct the following GKLS equation for $\hat{\varrho}_t \in \mathcal{D}(\mathbf{H})$,
\begin{widetext}
\begin{eqnarray}\label{EnlargedME}
    \frac{d\hat{\varrho}_t}{dt} &=& i[\hat{\varrho}_t, \hat{\mathcal{H}}] - \frac{1}{2} \sum_i [ \hat{\mathcal{L}}_i^\dagger\hat{\mathcal{L}}_i \hat{\varrho}_t + \hat{\varrho}_t \hat{\mathcal{L}}_i^\dagger\hat{\mathcal{L}}_i - 2 \hat{\mathcal{L}}_i\hat{\varrho}_t\hat{\mathcal{L}}_i^\dagger] - \frac{1}{2} [ \hat{\mathcal{L}}_D^\dagger\hat{\mathcal{L}}_D \hat{\varrho}_t + \hat{\varrho}_t \hat{\mathcal{L}}_D^\dagger\hat{\mathcal{L}}_D - 2 \hat{\mathcal{L}}_D\hat{\varrho}_t\hat{\mathcal{L}}_D^\dagger],
\end{eqnarray}
\end{widetext}
where $\hat{\mathcal{H}} = \begin{pmatrix} \hat{M} & 0 \\ 0 & 0 \end{pmatrix}$ is the Hamiltonian and $\hat{\mathcal{L}}_i = \begin{pmatrix} \hat{L}_i & 0 \\ 0 & 0 \end{pmatrix}$, in accordance with the notation in Eq.~(\ref{GenME}). Let us show that the decay property is governed by Eq.~(\ref{EnlargedME}) if the Lindblad operator $\hat{\mathcal{L}}_D$ has the form
\begin{equation}
 \hat{\mathcal{L}}_D = \begin{pmatrix} 0 & 0 \\ \hat{L}_D & 0 \end{pmatrix}.
\end{equation}
This choice means that $\hat{\mathcal{L}}_D$ represents a transition between the flavor $\mathbf{H}_M$ and the ``decay'' $\mathbf{H}_D$ subspaces of $\mathbf{H}$. In this way, if $\mathbf{H}_D$ has the same dimensions as $\mathbf{H}_M$, then we can decompose $\hat{L}_D$, which acts in $\mathbf{H}_M$, as
\begin{equation}
    \hat{L}_D = \sum_i \sqrt{\gamma_i} |f_i\rangle\langle M_i|,
\end{equation}
where $\gamma_i \geq 0$. By projecting the master equation~(\ref{EnlargedME}) back to $\mathbf{H}_M$ we obtain a master equation
\begin{widetext}
\begin{eqnarray}
\frac{d\hat{\rho}_t}{dt} &=& i[\hat{\rho}_t, \hat{M}] - \frac{1}{2} \sum_i [ \hat{L}_i^\dagger\hat{L}_i \hat{\rho}_t + \hat{\rho}_t \hat{L}_i^\dagger\hat{L}_i - 2 \hat{L}_i\hat{\rho}_t\hat{L}_i^\dagger] - \frac{1}{2}\{\hat{L}_D^\dagger \hat{L}_D, \hat{\rho}_t\} \label{ImDecayME}.
\end{eqnarray}
\end{widetext}
Comparing~(\ref{ImDecayME}) with the master equation~(\ref{GenME}), we see that they are identical if the decay operator is set to be
\begin{equation}
\hat{\Gamma} = \hat{L}_D^\dagger \hat{L}_D = \sum_i \gamma_i |M_i\rangle\langle M_i |, \label{DecayOp}
\end{equation}
so that $\gamma_i = \Gamma_i$ are simply the decay widths of the corresponding lifetime eigenstates $|M_i\rangle$ of the non-Hermitian Hamiltonian~(\ref{nonHermHam}). Hence, the choice~(\ref{DecayOp}) of the decay operator guarantees the consistency of the evolution of a $M^0$--$\bar{M}^0$ system induced by the master equation~(\ref{GenME}). 

\section{Collapse dynamics in $\mathbf{H}_M$ \label{collapsedyn}}
In a typical advanced dynamical reduction model, such as the QMUPL and CSL models discussed in Section~\ref{qmuplcsl}, the wave function collapse is considered as a continuous physical process with respect to the interaction between the quantum system and a randomly fluctuating (noise) field. Mathematically, this is achieved by a non-linear stochastic modification of the Schr\"{o}dinger equation\footnote{In this Section, we use the It\=o stochastic formalism, for details see Appendix~\ref{StochProp}.},
\begin{widetext}
\begin{eqnarray}\label{NonHermCollapseEq}
d|\psi\rangle_t = \Bigl[ -i\hat{H} dt + \sqrt{\lambda} \sum_i \Bigl(\hat{A}_i - R_{\hat{A}_i} \Bigr) dW_{i,t} - \frac{\lambda}{2} \Bigl[ \hat{A}_i^\dagger \hat{A}_i - 2 R_{\hat{A}_i} \hat{A}_i + R_{\hat{A}_i}^2 \Bigl] dt \Bigr] |\psi\rangle_t,
\end{eqnarray}
\end{widetext}
with
\begin{equation}
R_{\hat{A}_i} = \Bigl\langle \frac{\hat{A}_i^\dagger + \hat{A}_i}{2} \Bigr\rangle_t,
\end{equation}
where $\hat{H}$ governs the standard Schr\"{o}dinger part of the evolution, $\lambda \geq 0$ is the coupling constant of the collapse model which represents the localization rate, $W_{i,t}$ is a set of the Wiener processes, $\hat{A}_i$ are the corresponding collapse operators, and $\langle \hat{A}_i \rangle_t := \,{}_t\langle \psi| \hat{A}_i |\psi\rangle_t$ is the quantum mechanical expectation value.

The change of the Wiener process $W_{i,t}$ in time is represented by the mentioned above noise field $dW_{i,t}$, whose correlation function reads
\begin{equation}\label{CorrFunc}
    \mathbb{E}[dW_{i,t}dW_{j,t'}] = \delta_{ij}\delta(t-t'),
\end{equation}
where $\mathbb{E}$ denotes the noise average. This noise field is white, i.e., all its frequencies equally contribute to the collapse process. The nature of the noise field still remains an open question: one of the options is a physical field filling the whole space, hence, having presumably a cosmological nature (for example, the relic cosmic neutrino background would be a possible candidate for the noise field of the CSL model~\cite{Mishra2018}). Hence, there is a challenge to study the collapse models with the noise field whose properties differ from ones of a 'typical' white noise~\cite{AdlerBassi2007, AdlerBassi2008, BassiFerialdi2009a, BassiFerialdi2009b, FerialdiBassi2012}. In turn, the noise field
can depend on the direction of time\footnote{Mathematically, this dependence on the time direction is intimately connected to the asymmetry of the correlation function~(\ref{CorrFunc}) and, in turn, the definition of the stochastic integral. We discuss the mathematical aspects of the asymmetry of (\ref{CorrFunc}) in Appendix~\ref{StochProp}.}, which plays a crucial role in decay dynamics of a $M^0$--$\bar{M}^0$ system as we show later in this Section. 

The collapse operators $\hat{A}_i$ define the preferred basis, into which the state of the quantum system is reduced. A particular interest has the case of the self-adjoint collapse operators, which significantly simplifies the collapse equation~(\ref{NonHermCollapseEq}),
\begin{eqnarray}\label{CollapseEq}
\nonumber d|\psi\rangle_t &=& \Bigl[ -i\hat{H} dt + \sqrt{\lambda} \sum_i \Bigl(\hat{A}_i - \langle \hat{A}_i \rangle_t \Bigr) dW_{i,t} \\
&-& \frac{\lambda}{2} \sum_i (\hat{A}_i - \langle \hat{A}_i \rangle_t )^2 dt \Bigr] |\psi\rangle_t. \end{eqnarray}
For a neutral meson system, whose standard Schr\"{o}dinger evolution is governed by the phenomenological Hamiltonian~(\ref{nonHermHam}), the (flavor) $\mathbf{H}_M$-counterpart of the collapse dynamics can be governed by the self-adjoint collapse operator
\begin{equation}\label{FlavorCollapseOp}
\hat{A} = \sum_i \widetilde{m}_i |M_i\rangle\langle M_i|,
\end{equation}
where $\widetilde{m}_i$ is the mass ratio with respect to the corresponding mass $m_i$ and the reference mass $m_0$. The latter can be seen as a free parameter of a collapse model which relates (generally speaking, for ordinary matter) the mass ratio to an average number of constituents of the composite object and tunes the amplification mechanism of the collapse model~\cite{Pearle1994}. In a typical collapse model (in particular, the mass-proportional CSL model discussed in Section~\ref{qmuplcsl}), the amplification mechanism strengthens the collapse effect for a more massive object suggesting the definition $\widetilde{m}_i = \frac{m_i}{m_0}$ of the mass ratio\footnote{Neutral meson systems reveal opposite behavior: lighter particles decay faster, and the mass ratio $\frac{m_i}{m_0}$ decreases. Thus, as discussed in~\cite{SimonovPaper}, an inverted mass ratio $\widetilde{m}_i = \frac{m_0}{m_i}$ would be more reasonable for the particles lighter than the constituents of ordinary matter. However, in this paper, we imply that $\widetilde{m}_i$ represents the "usual" mass ratio, unless otherwise noted.}.

Accepting Eq.~(\ref{CollapseEq}) as the dynamical equation for the state $|\psi\rangle_t \in \mathbf{H}_M$ of a neutral meson system, it is possible to derive the master equation~(\ref{GenME}) for the corresponding density operator $\hat{\rho}_t = \mathbb{E}[|\psi\rangle_t\langle\psi|_t]$ from it. In this case, the role of the Lindblad generators $\hat{L}_i$ play the collapse operators weighted by the localization rate, so that $\hat{L}_i = \sqrt{\lambda} \hat{A}_i$. However, as we have discussed in Section~\ref{particles}, Eq.~(\ref{GenME}) is derived with respect to the non-Hermitian Hamiltonian $\hat{H}$. Hence, it does not guarantee itself the complete positivity of the resulting dynamics, which could produce no physically consistent state of the neutral meson system in this case. Therefore, we would desire to have a collapse model in $\mathbf{H}$ which induces the GKLS master equation~(\ref{EnlargedME}). This guarantees that Eq.~(\ref{CollapseEq}) is associated with the master equation~(\ref{ImDecayME}), which is a projection of Eq.~(\ref{EnlargedME}) onto $\mathbf{H}_M$ and, hence, induces a physically consistent dynamics of the $M^0$--$\bar{M}^0$ system. Let us build such collapse dynamics in $\mathbf{H}$. At first, we have to reproduce the self-adjoint collapse operator which induces the reduction in the mass basis due to~(\ref{FlavorCollapseOp}), namely, $\hat{\mathcal{A}} = \begin{pmatrix} \hat{A} & 0 \\ 0 & 0 \end{pmatrix}$. Another (non-Hermitian) collapse operator $\hat{\mathcal{B}} = \begin{pmatrix} 0 & 0 \\ \hat{B} & 0 \end{pmatrix}$ with 
\begin{equation}
\hat{B} = \sum_i \sqrt{\frac{\Gamma_i}{\lambda}} |f_i\rangle\langle M_i |
\end{equation}
triggers the decay of a neutral meson to the product states $|f_i\rangle$ with the corresponding decay widths $\Gamma_i$. With a Hermitian Hamiltonian $\hat{\mathcal{H}} = \begin{pmatrix} \hat{M} & 0 \\ 0 & 0 \end{pmatrix}$, the required dynamics can be governed by the following differential equation for a quantum state $|\Psi\rangle_t \in \mathbf{H}$,
\begin{widetext}
\begin{eqnarray}
    d|\Psi\rangle_t &=& \Biggl\{ -i\hat{\mathcal{H}} dt + \sqrt{\lambda} \Bigl[ (\hat{\mathcal{A}} - \langle\hat{\mathcal{A}} \rangle_t) dW_{t} + \Bigl(\hat{\mathcal{B}} - R_{\hat{\mathcal{B}}} \Bigr) dW^D_t \Bigr] - \frac{\lambda}{2} \Bigl[ (\hat{\mathcal{A}} - \langle \hat{\mathcal{A}} \rangle_t)^2 + \hat{\mathcal{B}}^\dagger \hat{\mathcal{B}} - 2 R_{\hat{\mathcal{B}}} \hat{\mathcal{B}} + R_{\hat{\mathcal{B}}}^2 \Bigl] dt\Biggr\} |\Psi\rangle_t, \label{EnlargedSDE}
\end{eqnarray}
\end{widetext}
with
\begin{equation}
R_{\hat{\mathcal{B}}} = \Bigl\langle \frac{\hat{\mathcal{B}}^\dagger + \hat{\mathcal{B}}}{2} \Bigr\rangle_t.
\end{equation}
It can be shown that Eq.~(\ref{EnlargedSDE}) is associated with the GKLS master equation~(\ref{EnlargedME}) for a density operator $\hat{\varrho}_t = \mathbb{E}[|\Psi\rangle_t \langle \Psi |_t]$, where the corresponding Lindblad generators are proportional to the collapse operators, $\hat{\mathcal{L}} = \sqrt{\lambda} \hat{\mathcal{A}}$ and $\hat{\mathcal{L}}_D = \sqrt{\lambda} \hat{\mathcal{B}}$.

Going back to the flavor space $\mathbf{H}_M$, we project the obtained collapse equation~(\ref{EnlargedSDE}) onto it. In this way, we find the following collapse equation for $|\psi\rangle_t \in \mathbf{H}_M$,
\begin{eqnarray}\label{ItoEqDecay}
    \nonumber d|\psi\rangle_t &=& \Bigl[ -i\hat{M} dt + \sqrt{\lambda} \Bigl(\hat{A} - \langle \hat{A} \rangle_t \Bigr) dW_{t} \\
&-& \frac{\lambda}{2} \Bigl( (\hat{A} - \langle \hat{A} \rangle_t )^2 + \hat{B}^\dagger \hat{B} \Bigr) dt \Bigr] |\psi\rangle_t,
\end{eqnarray}
which is associated with the master equation~(\ref{ImDecayME}). Comparing Eq.~(\ref{ItoEqDecay}) with the original collapse equation~(\ref{CollapseEq}), we see that there appears a drift term proportional to $\lambda \hat{B}^\dagger\hat{B}$ which plays the role of the non-Hermitian part $\hat{\Gamma}$ of the phenomenological Hamiltonian~(\ref{nonHermHam}) governing the decay property, as we would expect from the master equation~(\ref{EnlargedME}). However, when looking at Eqs.~(\ref{EnlargedSDE}) and~(\ref{ItoEqDecay}), the collapse mechanism seems to have no impact on the decay dynamics since its localization rate $\lambda$, in fact, does not show up in the terms associated with the decay, so that
\begin{eqnarray}
 \nonumber \mathbb{E}[d|||\psi\rangle_t||^2] &=& -\lambda \mathbb{E}[\langle \psi |_t
 \hat{B}^\dagger\hat{B}|\psi\rangle_t]dt \\
 &=& -\sum_i \Gamma_i \mathbb{E}[|\langle M_i|\psi\rangle_t|^2]dt \neq 0,
\end{eqnarray}
in accordance with the predictions of the effective Schr\"{o}dinger equation (\ref{nonHermSchr}) for the dynamics of neutral mesons. Indeed, the role of "localization rates" for the collapse operators $\hat{\mathcal{B}}$ and $\hat{B}$ play the decay widths $\Gamma_i$ inserted by hand, and we just reproduce the $\hat{\Gamma}$ operator in accordance with~(\ref{DecayOp}). However, it is possible to obtain spontaneous collapse dynamics which does not simply mimic the decay operator $\hat{\Gamma}$ but rather induces additional energy terms which play the role of the decay widths $\Gamma_i$ and, hence, recovers the decay dynamics of a $M^0$--$\bar{M}^0$ system.


Before to proceed, let us give a remark on the properties of dynamics governed by Eqs.~(\ref{ImDecayME}) and~(\ref{CollapseEq}). Despite the non-linearity of the quantum state equation~(\ref{CollapseEq}), which makes it difficult to solve, its physical predictions can be analyzed in a simple way by using a peculiar mathematical property of Eqs.~(\ref{ImDecayME}) and~(\ref{CollapseEq}). The master equation~(\ref{ImDecayME}) and, in turn, the physical predictions of the quantum state equation~(\ref{CollapseEq}) concerning the outcomes of a measurement (in particular, the transition probabilities~(\ref{ProbMS}) and~(\ref{ProbFS})) turn out to be invariant\footnote{This transformation introduced in~\cite{AdlerBassi2007} can be generalized to the transformation $
d|\psi\rangle_t = \Bigl[ -i\hat{H} dt + \sqrt{\lambda} \sum_i \Bigl(e^{i\varphi}\hat{A}_i - \xi\langle\hat{A}_i\rangle_t \Bigl) dW_{i,t} - \frac{\lambda}{2} \sum_i \Bigl(\hat{A}_i^2 - 2 e^{i\varphi} \xi^* \langle\hat{A}_i\rangle_t \hat{A}_i + (|\xi|^2 + i \omega) \langle\hat{A}_i\rangle_t^2 \Bigr) dt \Bigr] |\psi\rangle_t$ with additional parameters $\xi \in \mathbb{C}$ and $\omega \in \mathbb{R}$, which still leaves the master equation~(\ref{ImDecayME}) invariant.} under the phase transformation~\cite{AdlerBassi2007},
\begin{widetext}
\begin{equation}
d|\psi\rangle_t = \Bigl[ -i\hat{H} dt + \sqrt{\lambda} \sum_i \Bigl(e^{i\varphi}\hat{A}_i - \cos(\varphi)\langle\hat{A}_i\rangle_t \Bigl) dW_{i,t} - \frac{\lambda}{2} \sum_i \Bigl(\hat{A}_i^2 - 2 e^{i\varphi} \cos(\varphi) \langle\hat{A}_i\rangle_t \hat{A}_i + \cos^2(\varphi) \langle\hat{A}_i\rangle_t^2 \Bigr) dt \Bigr] |\psi\rangle_t,
\end{equation}
\end{widetext}
of Eq.~(\ref{CollapseEq}). In particular, the original collapse equation~(\ref{CollapseEq}) can be recovered by choosing $\varphi = 0$, whereas the choice $\varphi = \frac{\pi}{2}$ introduces an imaginary noise field and leads to a more simple Schr\"{o}dinger-like equation
\begin{equation} \label{ItoImEq}
d|\psi\rangle_t = [ -i\hat{H} dt + i \sqrt{\lambda} \sum_i \hat{A}_i dW_{i,t} - \frac{\lambda}{2} \sum_i \hat{A}_i^2 dt] |\psi\rangle_t.
\end{equation}
Applying the noise field transformation to Eq.~(\ref{ItoEqDecay}) we obtain the quantum state equation
\begin{equation} \label{ItoNonHermEq}
d|\psi\rangle_t = [ -i\hat{M} dt + i \sqrt{\lambda} \hat{A} dW_{t} - \frac{\lambda}{2} (\hat{A}^2 + \hat{B}^\dagger \hat{B}) dt] |\psi\rangle_t.
\end{equation}
In the absence of the decay contribution $\hat{B}^\dagger \hat{B}$, the quantum state equation~(\ref{ItoNonHermEq}) conserves the norm of $|\psi_t\rangle$ and, indeed, there is no built-in direction of time\footnote{Notice that the collapse models, generally speaking, do not necessarily need a built-in direction of time for the description of spontaneous collapse, which can seem an essentially time-asymmetric process~\cite{Bedingham2017, Bedingham2018}.}. As discussed in Appendix~\ref{StochProp}, this can be interpreted as the time-symmetry of the noise field, i.e., the action of the noise field in the bra-space (``out''-states) and ket-space (``in''-states) is symmetric. From the other hand, considering a class of more general collapse models with a time-asymmetric noise is more reasonable for the decay dynamics, which does not conserve the norm of $|\psi\rangle_t$. It generates a quantum state equation
\begin{equation} \label{ItoNonHermEqFamily}
d|\psi\rangle_t = [ -i\hat{M} dt + i \sqrt{\lambda} \hat{A} dW_{t} - \lambda \beta \hat{A}^2 dt] |\psi\rangle_t,
\end{equation}
with the parameter $\beta \in [0, 1]$ which describes this time-asymmetry\footnote{The choice $\beta = \frac{1}{2}$ gives a "usual" noise field with no dependence on the direction of time.} and can be seen as tuning of the coupling between the bra- and ket-spaces with respect to the noise field action~\cite{SimonovPaper}. Comparing Eq.~(\ref{ItoNonHermEqFamily}) with Eq.~(\ref{ItoNonHermEq}) we see that the decay operator $\hat{\Gamma}$ can be recovered by an operator induced by the collapse dynamics in $\mathbf{H}_M$,
\begin{equation}\label{DecayOpCollapse}
   \hat{\Gamma} \equiv \lambda \hat{B}^\dagger \hat{B} = -\lambda (1 - 2\beta) \hat{A}^2.
\end{equation}
Decomposing the collapse operators $\hat{A}$ and $\hat{B}$, we obtain
\begin{equation}\label{DecayWidth}
    \Gamma_i = -\lambda (1 - 2\beta)\widetilde{m}_i^2
\end{equation}
in the definition\footnote{The overall minus sign comes from our convention on the direction of time arrow, so that the non-negative decay widths correspond to $\beta \in [\frac{1}{2}, 1]$ in this case.} of the collapse operator $\hat{B}$. This implies that the spontaneous collapse can be an only source of the two distinct decay widths, and, in turn, the decay dynamics in $\mathbf{H}_M$ is governed by the collapse dynamics.

Summarizing, the non-Hermitian part of the Wigner-Weisskopf effective Hamiltonian~(\ref{nonHermHam}), which governs the decay property of neutral mesons, can be induced by spontaneous collapse dynamics in $\mathbf{H}_M$ with respect to the quantum state equation~(\ref{ItoNonHermEqFamily}) or, equivalently, master equation
\begin{eqnarray}
\nonumber \frac{d\hat{\rho}_t}{dt} &=& i[\hat{\rho}_t, \hat{M}] - \frac{\lambda}{2} [ \hat{A}^2 \hat{\rho}_t + \hat{\rho}_t \hat{A}^2 - 2 \hat{A}\hat{\rho}_t\hat{A}] \\
&+& \frac{\lambda}{2} (1 - 2\beta)\{\hat{A}^2, \hat{\rho}_t\}. \label{FamilyME}
\end{eqnarray}
In other words, a broader class of time-asymmetric collapse models can in principle gain the decay property of neutral mesons. As shown in~\cite{SimonovPaper, SimonovKLOE}, such a collapse model makes possible to predict the absolute masses $m_{H/L}$ of the lifetime eigenstates as well as the collapse parameters $\lambda$ and $\beta$ using the experimentally measured values of the decay constants $\Gamma_{H/L}$ and the difference of masses $\Delta m = m_H - m_L$. We discuss these interesting consequences of the spontaneous collapse dynamics in the following Section.

\section{QMUPL and CSL models \label{qmuplcsl}}
In the previous Section, we have obtained a class of collapse models which could explain the full dynamics of a $M^0$--$\bar{M}^0$ system in the flavor Hilbert space $\mathbf{H}_M$, where the oscillations take place. Now, we turn to the collapse models acting in the Hilbert space $L^2 (\mathbb{R}^d) \otimes \mathbf{H}_M$ with $d=1,2,3$, which combines the position and flavor spaces, and the collapse is assumed to the spatial part of the state of the system. For the sake of concreteness, we consider the QMUPL and the mass-proportional CSL models, which were widely analyzed in the context of flavor oscillations~\cite{Bahrami2013, Donadi2012, Donadi2013, SimonovLetter, SimonovPaper, SimonovKLOE}, and generalize them to the models with the time-asymmetric noise field. In this way, we analyze the quantum state equation
\begin{equation} \label{FinalEquation}
d|\psi\rangle_t = [ -i\hat{M} dt + i \sqrt{\lambda} \sum_i \hat{\mathbf{A}}_i dW_{i,t} - \lambda \beta \sum_i \hat{\mathbf{A}}_i^2 dt] |\psi\rangle_t,
\end{equation}
with the collapse operators 
\begin{eqnarray}
\hat{\mathbf{A}}^{QMUPL}_{i} &=& \hat{q}_i \otimes \hat{A}, \label{OperatorQMUPL} \\
\hat{\mathbf{A}}^{CSL}_{\mathbf{x}} &=& \hat{Q}(\mathbf{x}) \otimes \hat{A}, \label{OperatorCSL}
\end{eqnarray}
of the QMUPL and the mass-proportional CSL models, respectively. In the QMUPL model, $\hat{q}_i$ denotes the $i$-th coordinate operator in $L^2(\mathbb{R}^d)$, and the unique introduced collapse parameter is the collapse rate $\lambda \equiv \lambda_Q$ which has the units $[(m^2 \cdot s)^{-1}]$. In the mass-proportional CSL model,
\begin{equation}\label{PositionCSL}
\hat{Q}(\mathbf{x}) = \int \frac{d\mathbf{y}}{(2\pi r_C^2)^{\frac{d}{2}}} e^{-\frac{(\mathbf{x}-\mathbf{y})^2}{2r_C^2}} |\mathbf{y}\rangle\langle\mathbf{y}|
\end{equation}
is a continuous set of collapse operators acting in $L^2(\mathbb{R}^d)$ which are smeared by a Gaussian function\footnote{Originally, the CSL model is formulated within second quantization formalism, in order to describe a system of indistinguishable particles. For the sake of simplicity, we formulate it as a first-quantized model following~\cite{Tilloy2017}.}. In contrast to the QMUPL model, the mass-proportional CSL model introduces two natural constants, the collapse rate being traditionally denoted $\lambda \equiv \gamma$, which has the units $[m^d \cdot s^{-1}]$, and the coherence length $r_C$.

At accelerator facilities, one intensively studies the decay modes of neutral mesons. Observation of a decay mode is a passive measurement procedure, which allows the experimenter to determine the lifetime of a neutral meson or its flavor content (for example, strangeness in the case of K-mesons). Hence, in standard quantum mechanics, it leads to the corresponding lifetime and flavor measurement probabilities given by the transition probabilities~(\ref{ProbMS}) and~(\ref{ProbFS}). With included contribution from spontaneous collapse, these probabilities can be defined through
\begin{equation}\label{ProbQS}
    P_{\mathrm{in} \rightarrow \mathrm{out}}(t) = |\langle\psi_{\mathrm{out}}|\hat{U}(t)|\psi_{\mathrm{in}}\rangle|^2
\end{equation}
for $|\psi_{\mathrm{in},\mathrm{out}}\rangle \in \mathbf{H}_M$, where $\hat{U}(t)$ is the evolution operator due to the quantum state equation~(\ref{FinalEquation}). The detailed perturbative calculations of (\ref{ProbQS}) for the lifetime eigenstates $|M_i\rangle$ and the flavor states $|M^0/\bar{M}^0\rangle$ are already known in literature~\cite{Donadi2013, SimonovLetter, SimonovPaper, SimonovKLOE}, hence, we choose a simpler procedure and stick to the class of master equations in accordance with~(\ref{FamilyME}), namely,
\begin{eqnarray}\label{MasterEqQMUPLCSL}
\nonumber \frac{d\hat{\rho}_t}{dt} &=& i[\hat{\rho}_t, \hat{M}] - \frac{\lambda}{2} \sum_i [ \hat{\mathbf{A}}_i^2 \hat{\rho}_t + \hat{\rho}_t \hat{\mathbf{A}}_i^2 - 2 \hat{\mathbf{A}}_i\hat{\rho}_t\hat{\mathbf{A}}_i] \\
&+& \frac{\lambda}{2} (1 - 2\beta) \sum_i \{\hat{\mathbf{A}}_i^2, \hat{\rho}_t\}, \label{FinalME}
\end{eqnarray}
with the collapse operators~(\ref{OperatorQMUPL}) or~(\ref{OperatorCSL}). Solving Eq.~(\ref{FinalME}), we can derive the required transition probabilities~(\ref{ProbQS}) which now read
\begin{equation}\label{ProbDM}
P_{\mathrm{in} \rightarrow \mathrm{out}}(t) = \operatorname{Tr}[ (\mathbb{1} \otimes |\psi_{\mathrm{out}}\rangle\langle \psi_{\mathrm{out}}|) \hat{\rho}_t],
\end{equation}
where the initial state is chosen as $\hat{\rho}_{0} = |\alpha\rangle\langle\alpha| \otimes |\psi_{\mathrm{in}}\rangle\langle\psi_{\mathrm{in}}|$ with $|\alpha\rangle$ representing a Gaussian wave packet of width $\sqrt{\alpha}$.

Following the calculations done in the Appendix~\ref{QMUPLCSLComputations}, we find that the QMUPL collapse effect is not exponential, while the CSL model recovers the exponential effect, in accordance with~\cite{SimonovPaper}. In particular, the transition probabilities for the lifetime eigenstates read
\begin{eqnarray}
    P_{M_i \rightarrow M_j}^{QMUPL}(t) &=& \Bigl( 1 - \lambda_Q \alpha (1-2\beta)\widetilde{m}_i^2 t\Bigr)^{-\frac{d}{2}} \delta_{ij}, \\
    P_{M_i \rightarrow M_j}^{CSL}(t) &=& e^{-\frac{\gamma}{(\sqrt{4\pi} r_C)^d} (2\beta - 1) \widetilde{m}_i^2 t} \delta_{ij},
\end{eqnarray}
whereas the transition probabilities for the flavor states, which reveal the particle oscillations, read
\begin{widetext}
\begin{eqnarray}
    \label{ProbQ} P_{M^0 \rightarrow M^0/\bar{M}^0}^{QMUPL}(t) &=& \frac{1}{4}\Biggl\{ \sum_i \Bigl( 1 - \lambda_Q \alpha (1-2\beta) \widetilde{m}_i^2 t\Bigr)^{-\frac{d}{2}} \pm \frac{2\cos[t \Delta m]}{\Bigl( 1 - \frac{\lambda_Q \alpha}{2} \Bigl((1 - 2\beta) \sum_i \widetilde{m}_i^2 - (\Delta \widetilde{m})^2 \Bigr) t\Bigr)^{\frac{d}{2}}} \Biggr\}, \\
    \label{ProbC} P_{M^0 \rightarrow M^0/\bar{M}^0}^{CSL}(t) &=& \frac{1}{4}\Bigl\{ \sum_i e^{-\frac{\gamma}{(\sqrt{4\pi}r_C)^d}(2\beta - 1) \widetilde{m}_i^2 t} \pm 2 e^{-\frac{1}{2}\frac{\gamma}{(\sqrt{4\pi}r_C)^d} ((2\beta - 1) \sum_i \widetilde{m}_i^2 + (\Delta \widetilde{m})^2 ) t} \cos[t \Delta m] \Bigr\}.
\end{eqnarray}
\end{widetext}
In particular, the oscillations of K- and B-mesons were intensively studied through their (semi-)leptonic decays
\begin{eqnarray}
\nonumber K^0 &\rightarrow& \pi^- e^+ \nu_e, \\
\nonumber \bar{K}^0 &\rightarrow& \pi^+ e^- \bar{\nu}_e, \\
\nonumber B^0 &\rightarrow& D^- \mu^+ \nu_\mu, \\
\nonumber \bar{B}^0 &\rightarrow& D^+ \mu^- \bar{\nu}_\mu,
\end{eqnarray}
at accelerator facilities in the CPLEAR and the BaBar and Belle experiments, respectively. In these experiments, detecting a lepton of a definite charge uniquely identifies the flavor of the decayed meson. In this way, a direct measure of the relative flavor components of the state of a neutral meson is provided by the relative decay rates. The corresponding decay rates can be associated with the transition probabilities for the flavor states up to a normalization factor. Hence, the asymmetry term, which is a combination of the flavor transition probabilities,
\begin{equation}
\mathbb{A}(t) = \frac{P_{M^0 \rightarrow M^0}(t) - P_{M^0 \rightarrow \bar{M}^0}(t)}{P_{M^0 \rightarrow M^0}(t) + P_{M^0 \rightarrow \bar{M}^0}(t)}    
\end{equation}
plays a significant role by removing the overall normalization factor and, hence, cancelling potential systematic biases~\cite{Thomson2013}. For the standard quantum mechanical transition probabilities~(\ref{ProbFS}), it reads
\begin{equation}\label{AsymQM}
\mathbb{A}(t) = \frac{\cos[t \Delta m]}{\cosh\Bigl[\frac{\Delta\Gamma}{2} t\Bigr]},
\end{equation}
where $\Delta\Gamma = \Gamma_L - \Gamma_H$. Plugging in the transition probabilities~(\ref{ProbQ}) and~(\ref{ProbC}), we include the effects of the QMUPL and the CSL models into the asymmetry term and obtain
\begin{widetext}
\begin{eqnarray}
    \mathbb{A}^{QMUPL}(t) &=& 2\cos[t \Delta m] \Biggl[ \Biggl( 1 - \frac{1}{2}\frac{\lambda_Q\alpha ((1-2\beta) \Delta \widetilde{m}^2 - (\Delta \widetilde{m})^2) t}{1 - \lambda_Q \alpha (1-2\beta) \widetilde{m}_L^2 t} \Biggl)^{\frac{d}{2}} + \Biggl( 1 + \frac{1}{2}\frac{\lambda_Q \alpha ((1-2\beta) \Delta \widetilde{m}^2  + (\Delta \widetilde{m})^2) t}{1 - \lambda_Q\alpha (1-2\beta) \widetilde{m}_H^2 t} \Biggl)^{\frac{d}{2}} \Biggr]^{-1}, \label{AsymQMUPL} \\
    \mathbb{A}^{CSL}(t) &=& \frac{\cos[t \Delta m]}{\cosh\Bigl[\frac{\gamma}{(\sqrt{4\pi}r_C)^d}(\beta - \frac{1}{2}) \Delta \widetilde{m}^2 t\Bigr]} \; e^{-\frac{\gamma}{(\sqrt{4\pi}r_C)^d} \frac{(\Delta \widetilde{m})^2 }{2} t}, \label{AsymCSL}
\end{eqnarray}
\end{widetext}
where $\Delta \widetilde{m}^2 = \widetilde{m}_H^2 - \widetilde{m}_L^2$. 
\begin{center}
\begin{figure}[h!]
\centering
\includegraphics[width=\columnwidth]{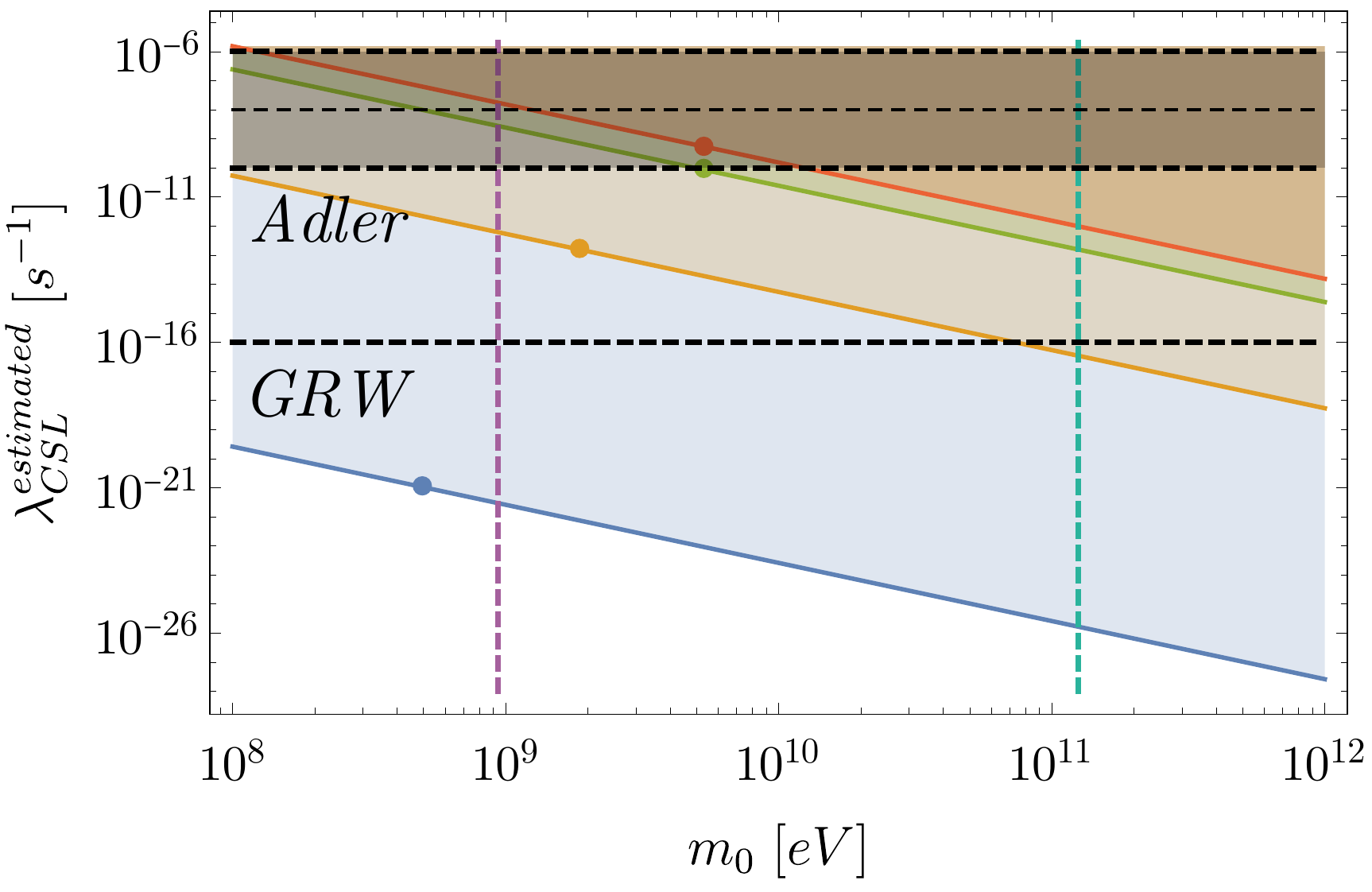}
\caption{Lower bounds for the estimated value of the collapse rate $\lambda_{CSL}$ via Eq.~(\ref{CollapseRateDecay}) as a function of the reference mass $m_0$. The blue, orange, green, and red areas correspond to the values of $\lambda_{CSL}^{estimated}$ given by K-, D-, B-, and B$_s$-mesons, respectively. The colored points refer to the rest masses of the corresponding mesons, the purple line is given for the nucleon mass, and the cyan line corresponds to the rest mass of the Higgs boson. There are plotted the theoretical values of the collapse rate $\lambda_{CSL} = 10^{-16}$s$^{-1}$ proposed by Ghirardi, Rimini, and Weber~\cite{GRW} and $\lambda_{CSL} = 10^{-8\pm 2}$s$^{-1}$ proposed by Adler~\cite{Adler2007} for $r_C = 10^{-7}$m. The inverted mass ratio $\widetilde{m}_i = \frac{m_0}{m_i}$ is assumed in accordance with~\cite{SimonovPaper}.}
\label{CollapseRateFigure}
\end{figure}
\end{center}
The asymmetry term~(\ref{AsymCSL}) with the included CSL collapse dynamics reveals two disentangled effects on the transition probabilities for neutral mesons. At first, there is exponential damping of the interference term, which depends on the mass difference $\Delta m$. Secondly, comparing~(\ref{AsymCSL}) with the standard quantum mechanical asymmetry term~(\ref{AsymQM}), which results from the Schr\"{o}dinger equation with the Wigner-Weisskopf effective Hamiltonian~(\ref{nonHermHam}), we find the induced exponential decay dynamics with the corresponding decay widths
\begin{equation}
\Gamma_i = -\frac{\gamma}{(\sqrt{4\pi} r_C)^d}(1 - 2\beta) \widetilde{m}_i^2,
\end{equation}
with respect to the phenomenological decay operator $\hat{\Gamma}$
\begin{equation}
\hat{\Gamma} = -\frac{\gamma}{(\sqrt{4\pi} r_C)^d} (1-2\beta)\hat{A}^2,
\end{equation}
as we would expect from~(\ref{DecayWidth}). The decay widths appear to depend on the absolute values $m_{H/L}$ of masses of neutral mesons, which play no role in standard quantum mechanics. This fact leads to two important consequences. On the one hand, for $\beta \neq \frac{1}{2}$, it allows to calculate the absolute masses in terms of experimentally measured quantities~\cite{SimonovPaper}, namely, decay widths $\Gamma_{H/L}$ and mass difference $\Delta m$ via the quadratic equation
\begin{eqnarray}\label{MassEq}
\frac{2\Delta\Gamma}{\Delta\Gamma \pm 2\Gamma} m_L^2 + 2(\Delta m) m_L + (\Delta m)^2 = 0,
\end{eqnarray}
where the upper sign corresponds to the normal mass ratio while the lower sign fixes the inverted mass ratio, which are discussed in Section~\ref{collapsedyn}. On the other hand, it is possible to estimate the collapse rate
\begin{equation}
\lambda_{CSL} \equiv \frac{\gamma}{(\sqrt{4\pi}r_C)^d} = \frac{\Gamma_i}{(2\beta - 1) \widetilde{m}_i^2}
\end{equation}
from~(\ref{DecayWidth}). In turn, the maximal value of the time-asymmetry $\beta = 1$ establishes a lower bound for the estimated collapse rate, which in terms of experimentally measured quantities reads
\begin{equation}
\lambda_{CSL}^{estimated} \geq \Biggl(\frac{\Delta m}{m_0\Bigl(\sqrt{\Gamma_L^{\pm 1}} - \sqrt{\Gamma_H^{\pm 1}}\Bigr)}\Biggr)^{\mp 2}, \label{CollapseRateDecay}
\end{equation}
where, as in~(\ref{MassEq}), the upper sign corresponds to the normal mass ratio while the lower sign fixes the inverted mass ratio. On Fig.~\ref{CollapseRateFigure}, the estimated values of $\lambda_{CSL}$ in dependence of the reference mass $m_0$ are plotted.

In the case of the QMUPL model, the neutral meson time evolution is still affected by these two effects, which, however, do not have exponential character (rather algebraic), so that they are in principle observable. Hence, the QMUPL and the mass-proportional CSL models offer distinguishable and non-equivalent dynamics for a $M^0$--$\bar{M}^0$ system because of the specific link between position and flavor Hilbert spaces. In this way, the mentioned quantities actively studied in the realm of particle physics crucially depend on the coupling between the particles and the noise field introduced by a collapse model and, thus, can give insight into the physics of the collapse mechanism.

\section{Conclusions}
Spontaneous collapse models propose a solution to the measurement problem in quantum mechanics by considering a collapse of a wave function as a physical process with respect to the interaction between the system and a (classical) noise field. These models predict interesting effects for the systems at higher energies. In particular, being applied to a flavor oscillating system, the QMUPL, and the mass-proportional CSL models, the two most popular spontaneous collapse models, predict damping of the oscillations. This distinguishes them, for example, from the semi-classical gravity approach equipped with the Schr\"{o}dinger-Newton equation~\cite{Diosi1984, Penrose1998, Bahrami2014}, which affects a neutral meson system by shifting the mass difference and, thus, changing the frequency of the flavor oscillations~\cite{Grossardt2015}.

Recently, it was shown that, in the spontaneous collapse models with a white noise field, its time (a)symmetry can lead to a non-trivial contribution to the quantities, which are usually studied at accelerator facilities~\cite{SimonovLetter, SimonovPaper, SimonovKLOE}. Indeed, collapse dynamics induced by the interaction of a neutral meson system with an time-asymmetric noise field results in a dependence on its absolute masses which does not appear in standard quantum mechanics. With respect to this contribution, the QMUPL model provides a non-exponential effect on the time evolution of a $M^0$--$\bar{M}^0$ system which is in principle observable. In turn, the mass-proportional CSL model can explain the decay property of a $M^0$--$\bar{M}^0$ system and recover its decay dynamics.

In our discussion, we have revisited the effects of the spontaneous collapse on neutral meson systems~\cite{SimonovLetter, SimonovPaper, SimonovKLOE}, and studied their connection with the properties of the underlying noise field. We have focused on the systems in the meson sector which provide the mixing of particles and antiparticles and are well suited for a non-relativistic quantum mechanical treatment since, in this case, the field-theoretical effect is negligible~\cite{Capolupo2004}. In this way, we can apply the non-relativistic collapse models such as the QMUPL and the CSL models in a consistent way. In principle, the results of this paper can be extended to any decaying mixed particles that allow for a non-relativistic treatment. However, there is a great challenge to construct a spontaneous collapse mechanism in relativistic QFTs~\cite{Nicrosini2003, Tumulka2006, Bedingham2010, Bedingham2011, Pearle2015, Tumulka2020}, and a closer look into the systems at higher energies would help to move forward the relativistic dynamical reduction program. One of the possible first steps towards it would be to analyze the effect of a non-white noise on a flavor oscillating system, which can be related to a field of a cosmological nature and is more physical than the white noise: the latter with a Lorenz-invariant correlation function produces infinite energy per unit time and unit volume~\cite{Pearle2015, Bassi2012}.

We have found that the decay property incorporated as a non-Hermitian part of a phenomenological Hamiltonian can result from collapse dynamics in an enlarged Hilbert space. Such a space includes the states which correspond to the decay products and is used to construct a Gorini--Kossakowski--Lindblad--Sudarshan (GKLS) equation, which allows to represent a decaying system as open and incorporate the particle decay as a Lindblad operator. In the following step, we have questioned whether spontaneous collapse dynamics can be the only source of the decay in the dynamics of a flavor oscillating system. Our analysis has shown that a broader class of collapse models with an asymmetric white noise field generates an extra term in the master equation, which induces the decay dynamics and depends on the absolute masses of neutral mesons. Analyzing the obtained master equation, we have confirmed the result of~\cite{SimonovPaper} for the generalized CSL model with an asymmetric noise field, whose collapse dynamics can fully describe the exponential decay behavior of a $M^0$--$\bar{M}^0$ system. In this way, we can conclude that the decay dynamics in the flavor oscillating systems can be described via the spontaneous collapse mechanism.

\vspace{0.5cm}
\textbf{Acknowledgements:}
K.S. acknowledges the Austrian Science Fund (FWF-P30821) and thanks Mathias Beiglb\"{o}ck (University of Vienna), Raffaele Del Grande (INFN-LNF), Sandro Donadi (Frankfurt Institute for Advanced Studies), Michael Kunzinger (University of Vienna), Eduard Nigsch (University of Vienna), and Andrea Smirne (Universit\`{a} degli Studi di Milano) for the fruitful discussions.

\appendix

\section{Stochastic formalisms and asymmetry of the noise field\label{StochProp}}
In stochastic calculus, one of the key objects of study is a stochastic differential equation, which describes the change of a stochastic process $X_t$ in time, 
\begin{equation}
 dX_t = f(X_t, t) dW_t + g(X_t, t) dt,
\end{equation}
where $f(X_t, t)$ is the so-called diffusion term, $g(X_t, t)$ is so-called drift term, and $W_t$ is a Wiener process with $W_0 = 0$, which represents the Brownian motion (in turn, $dW_t$ represents the white noise) and has the following important properties~\cite{Arnold1974},
\begin{itemize}
 \item it is normally distributed with the density $p(W_t) = \frac{1}{\sqrt{2\pi t}} e^{-\frac{W_t^2}{2t}}$,
 \item it has independent and stationary increments, so that $W_t - W_s$ depends only on $t-s$.
\end{itemize}
In order to solve such an equation, we have to define a stochastic integral $\int f dW_t$, which we desire to construct as a Riemann-Stieltjes integral. Focusing on the simplest case $f = W_t$, we define the stochastic integral as a mean-squared limit of the Riemann sums,
\begin{equation}\label{StochInt}
 \int_0^t W_t dW_t = \underset{n\rightarrow\infty}{\operatorname{ms-lim}} \sum_{i=1}^{n} W(t_i^*) [W(t_i) - W(t_{i-1})],
\end{equation}
where $\operatorname{ms-lim}$ denotes the mean-squared limit. In contrast to the usual Riemann integral, the value of the constructed stochastic integral~(\ref{StochInt}) depends on the choice of the intermediate points $t_i^*$, which lie in the intervals $[t_{i-1}, t_i]$, respectively. A particular choice of $t_i^*$ defines a formalism of stochastic calculus with its specific rules and properties. Generally speaking, there is a continuous family of stochastic formalisms with a corresponding choice of $t_i^*$. The two formalisms widely discussed in literature are the It\=o formalism, which corresponds to the left point choice $t_i^* = t_{i-1}$, and the Stratonovich formalism, which corresponds to the middle point choice $t_i^* = \frac{t_{i-1} + t_i}{2}$~\cite{Mannella, DePelo}. The It\=o formalism is widely used in financial mathematics since the integral, in this case, is a martingale, i.e., given only its history up to time $t_0$, the expectation value of the It\=o integral at any $t > t_0$ is simply its value at $t_0$~\cite{Roberts}. However, the rules of the It\=o calculus, generally speaking, do not coincide with those of the classical calculus: for example, the It\=o differentiation rule differs from the classical Leibniz's product rule and is given by
\begin{equation}
 d(X_t \cdot Y_t) = dX_t \cdot Y_t + X_t \cdot dY_t + \mathbb{E}[dX_t \cdot dY_t],
\end{equation}
where $\mathbb{E}$ denotes the stochastic average. The Stratonovich formalism, which is preferred in physics, operates with the usual rules of calculus.

In turn, the point $t_i^*$ can be connected to the value of the Heaviside function in zero by~\cite{Mannella, LauLubensky}
\begin{equation}\label{HeavisidePoint}
 t_i^* = t_{i-1} + \theta(0) (t_i - t_{i-1}),
\end{equation}
so that the It\=o formalism fixes $\theta(0) = 0$ and the Stratonovich formalism fixes $\theta(0) = \frac{1}{2}$. Typically, the Heaviside function appears in integrals of the correlation function $\mathbb{E}[dW_t dW_{t'}] = \delta(t-t')$ of the white noise since
\begin{equation}
    \int_{-\infty}^t \delta(t') dt' = \theta(t),
\end{equation}
and its value in zero $\theta(0)$, generally speaking, belongs to the interval $\theta(0) \in [0, 1]$ and is not well-defined. However, once the stochastic formalism is chosen, $\theta(0)$ is fixed due to Eq.~(\ref{HeavisidePoint}). This property can be used for an easy switch between different formalisms, which is governed by the formula
\begin{eqnarray}
    \nonumber f(X_t, t) \circ_\beta dW_t &=& f(X_t, t) \circ_{\beta'} dW_t \\
    &+& (\beta - \beta') \frac{\partial f (X_t, t)}{\partial X_t} f (X_t, t) dt,
\end{eqnarray}
where $\circ_\beta$ and $\circ_{\beta'}$ denote the product in the formalisms with $\theta(0) = \beta$ and $\theta(0) = \beta'$, respectively. In particular, the It\=o and Stratonovich conventions are related in the following way\footnote{Traditionally, the Strantonovich product is denoted by $\circ$, whereas the product sign is omitted for the It\=o formalism.},
\begin{eqnarray}
    \nonumber f(X_t, t) dW_t &=& f(X_t, t) \circ dW_t \\
    &-& \frac{1}{2} \frac{\partial f (X_t, t)}{\partial X_t} f (X_t, t) dt.
\end{eqnarray}
Except for the Stratonovich value $\theta_S(0)=\frac{1}{2}$, the white noise $dW_t$ as well as its correlation function $\delta_{\theta(0)}(t)$ are asymmetric with respect to the chosen value of $\theta(0)$, so that
\begin{equation}
 \int_0^t \delta_{\theta(0)}(t') dt' - \int^0_{-t} \delta_{\theta(0)}(t') dt' = 1 - 2\theta(0). 
\end{equation}
This asymmetry can be understood better by considering an approximation of $\delta_{\theta(0)}(t)$ by the asymmetric Laplace distribution,
\begin{equation}
 \delta_{\varkappa, \varepsilon}(t) = \frac{1}{\varepsilon} \frac{1}{\varkappa + \frac{1}{\varkappa}} e^{-\frac{|t|}{\varepsilon} \varkappa^{\operatorname{sign}(t)}},
\end{equation}
where $\varepsilon \ll 1$, and $\varkappa \in [0, \infty]$ is the parameter of asymmetry of the distribution. After some algebra, one can derive the following identity,
\begin{equation}
\theta(0) = \frac{\varkappa^2}{1 + \varkappa^2},
\end{equation}
which illustrates the connection between $\theta(0)$ and the asymmetry of the correlation function of the noise field characterized by the parameter $\varkappa$.

In accordance with the discussed rules of stochastic calculus, we can rewrite Eq.~(\ref{ItoNonHermEq}) in the Stratonovich formalism,
\begin{equation}\label{StratoImEqDecay}
d|\psi\rangle_t = \Bigl[ -i\hat{M} dt + i \sqrt{\lambda} \hat{A} \circ dW_t - \frac{\lambda}{2}\hat{B}^\dagger\hat{B} dt\Bigr] |\psi\rangle_t.
\end{equation}
The symmetric choice $\theta_S(0) = \frac{1}{2}$ (hence, time-symmetric noise field) in the Stratonovich formalism makes it clear that Eq.~(\ref{StratoImEqDecay}) can be interpreted as a Schr\"{o}dinger equation with the effective Hamiltonian~(\ref{nonHermHam}) with the decay operator $\hat{\Gamma} = \lambda\hat{B}^\dagger\hat{B}$, and a random potential. 

In the Stratonovich formalism, there is no built-in direction of time, which appears for other choices of the stochastic formalism and, hence, asymmetric $\theta(0)$. Let us change the formalism by shifting $\theta(0)$ from its Stratonovich value $\theta_S(0)$ to some another value $\beta$. This leads to the equation
\begin{eqnarray}
\nonumber d|\psi\rangle_t &=& [ -i\hat{M} dt + i \sqrt{\lambda} \hat{A} \circ_{\beta} dW_t \\
&-& \frac{\lambda}{2}\Bigr(\hat{B}^\dagger\hat{B} + (1-2\beta)\hat{A}^2\Bigl) dt] |\psi\rangle_t,
\end{eqnarray}
Assuming that the collapse dynamics in $\mathbf{H}_M$ is responsible for the decay property of a neutral meson system, i.e., plugging in the decay widths~(\ref{DecayWidth}), we finish with the stochastic Schr\"{o}dinger equation
\begin{equation}\label{ImEqDecay}
d|\psi\rangle_t = [ -i\hat{M} dt + i \sqrt{\lambda} \hat{A} \circ_{\beta} dW_t] |\psi\rangle_t,
\end{equation}
with, generally speaking, a time-asymmetric noise field $dW_t$ with respect to the chosen value of $\theta(0) = \beta$. In turn, Eq.~(\ref{ImEqDecay}) reads in Stratonovich formalism



\begin{equation}\label{ImEqDecayIto}
d|\psi\rangle_t = [ -i\hat{M} dt + i \sqrt{\lambda} \hat{A} \circ dW_t + \frac{\lambda}{2} (1-2\beta)\hat{A}^2 dt] |\psi\rangle_t.
\end{equation}
The equivalence of Eq.~(\ref{ImEqDecay}) and Eq.~(\ref{ImEqDecayIto}) means that the decay property can be gained by a collapse model with the underlying time-asymmetric noise field\footnote{Both Eq.~(\ref{ImEqDecay}) and Eq.~(\ref{ImEqDecayIto}) can be approximated, due to the Wong-Zaka\"i theorem~\cite{WongZakai}, by the same ordinary differential equation \[ i\frac{d}{dt}|\psi\rangle_t = [ \hat{M} - \sqrt{\lambda} \hat{A} \dot{W}_t^\varepsilon + i\frac{\lambda}{2} (1-2\beta)\hat{A}^2 ] |\psi\rangle_t, \] with a regularized noise $\dot{W}_t^\varepsilon = \int_\mathbb{R} \delta_\varepsilon(t - u) dW_u$ with a mollifier $\delta_\varepsilon$ converging to the delta function, so that its solutions converge to those of~(\ref{ImEqDecay}).}. Written in the It\=o formalism they correspond to Eq.~(\ref{ItoNonHermEqFamily}). Hence, the value of Heaviside function in zero $\theta(0)$ plays role of a natural constant introduced by such collapse model (along with $\lambda$ and $r_C$) and, in turn, provides physical sense to the choice of the stochastic formalism and the It\=o/Stratonovich dilemma. As mentioned in Section~\ref{collapsedyn}, it tunes the coupling between the bra- and ket-spaces concerning the action of the noise field and, in turn, sets the decay widths of the mixed particles.


\section{Computations of the transition probabilities for the QMUPL and the mass-proportional CSL models}\label{QMUPLCSLComputations}

Plugging the QMUPL and CSL collapse operators (\ref{OperatorQMUPL}) and (\ref{OperatorCSL}) into the master equation~(\ref{MasterEqQMUPLCSL}) we obtain
\begin{widetext}
\begin{eqnarray}
\frac{d\hat{\rho}_t}{dt} &=& i[\hat{\rho}_t, \hat{M}] - \lambda_Q \sum_i \Bigl[ \beta \{ \hat{q}_i^2 \otimes \hat{A}^2, \hat{\rho}_t \} - (\hat{q}_i \otimes \hat{A}) \hat{\rho}_t (\hat{q}_i \otimes \hat{A}) \Bigr], \label{MEQMUPL} \\
    \frac{d\hat{\rho}_t}{dt} &=& i[\hat{\rho}_t, \hat{M}] - \gamma\int d\mathbf{x} \; \Bigl[ \beta \{ \hat{Q}^2(\mathbf{x}) \otimes \hat{A}^2, \hat{\rho}_t \} -  (\hat{Q}(\mathbf{x}) \otimes \hat{A}) \hat{\rho}_t (\hat{Q}(\mathbf{x}) \otimes \hat{A}) \Bigl]. \label{MECSL}
\end{eqnarray}
\end{widetext}
These master equations can be used to calculate the required transition probabilities~(\ref{ProbDM}) in a simple way. At first, following~\cite{Tilloy2017}, we decompose the density operator $\hat{\rho}_t$ in the position and mass bases,
\begin{equation}
\hat{\rho}_t = \sum_{i, j = H, L} \int\int d\mathbf{x}d\mathbf{y} \rho_t^{ij}(\mathbf{x},\mathbf{y}) |\mathbf{x}\rangle\langle\mathbf{y}| \otimes |M_i\rangle\langle M_j|.
\end{equation}
This allows to rewrite the master equations~(\ref{MEQMUPL}) and~(\ref{MECSL}) for the QMUPL and CSL collapse dynamics, respectively, in terms of the matrix elements of the density operator $\hat{\rho}_t$
\begin{widetext}
\begin{eqnarray}
\frac{d}{dt} \rho_t^{ij}(\mathbf{x},\mathbf{y}) &=& \Bigl[ -i(m_i - m_j) - \frac{\lambda_Q}{2} \Bigl( |\widetilde{m}_i \mathbf{x} - \widetilde{m}_j \mathbf{y}|^2 - (1-2\beta) (\widetilde{m}_i^2 \mathbf{x}^2 + \widetilde{m}_j^2 \mathbf{y}^2) \Bigr) \Bigr] \rho_t^{ij}(\mathbf{x},\mathbf{y}), \\
\frac{d}{dt} \rho_t^{ij}(\mathbf{x},\mathbf{y}) &=& \Bigl[ -i(m_i - m_j) - \gamma \Bigl( \beta (\widetilde{m}_i^2 + \widetilde{m}_j^2) (g * g)(0) - \widetilde{m}_i \widetilde{m}_j (g*g)(\mathbf{x} - \mathbf{y}) \Bigr) \Bigr] \rho_t^{ij}(\mathbf{x},\mathbf{y}),
\end{eqnarray}
\end{widetext}
where 
\begin{equation}
    (g*g)(\mathbf{x}) = \int d\mathbf{y} g(\mathbf{y}) g(\mathbf{x}-\mathbf{y})
\end{equation}
is the convolution of two Gaussian smearing functions
\begin{equation}
    g(\mathbf{x} - \mathbf{y}) = \frac{1}{(2\pi r_C^2)^{\frac{d}{2}}} e^{-\frac{(\mathbf{x}-\mathbf{y})^2}{2r_C^2}}
\end{equation}
of the CSL collapse operator~(\ref{PositionCSL}). Solving these equations, we obtain
\begin{widetext}
\begin{eqnarray}
\rho_t^{ij}(\mathbf{x},\mathbf{y}) &=& e^{-i(m_i - m_j)t - \frac{\lambda_Q}{2}(|\widetilde{m}_i \mathbf{x} - \widetilde{m}_j \mathbf{y}|^2 - (1-2\beta) (\widetilde{m}_i^2 \mathbf{x}^2 + \widetilde{m}_j^2 \mathbf{y}^2)) t} \rho_0^{ij}(\mathbf{x},\mathbf{y}), \\
\rho_t^{ij}(\mathbf{x},\mathbf{y}) &=& e^{-i(m_i - m_j)t - \gamma (\beta(\widetilde{m}_i^2 + \widetilde{m}_j^2) (g * g)(0) - \widetilde{m}_i \widetilde{m}_j (g*g)(\mathbf{x} - \mathbf{y})) t} \rho_0^{ij}(\mathbf{x},\mathbf{y}).
\end{eqnarray}
\end{widetext}
Before to plug these solutions into~(\ref{ProbDM}), we have to calculate the partial trace of $\hat{\rho}_t$ over the position degrees of freedom, i.e., take the following integrals
\begin{widetext}
\begin{eqnarray}
\int d\mathbf{x} \;\rho_t^{ij}(\mathbf{x},\mathbf{x}) &=& \frac{e^{-i(m_i - m_j)t}}{\Bigl( 1 + \frac{\lambda_Q \alpha}{2}\Bigl( (\widetilde{m}_i - \widetilde{m}_j)^2 - (1-2\beta)(\widetilde{m}_i^2 + \widetilde{m}_j^2)\Bigr) t\Bigr)^{\frac{d}{2}}}, \\
\int d\mathbf{x} \;\rho_t^{ij}(\mathbf{x},\mathbf{x}) &=& e^{-i(m_i - m_j)t} e^{-\frac{1}{2} \frac{\gamma}{(\sqrt{4\pi}r_C)^d}((\widetilde{m}_i - \widetilde{m}_j)^2 - (1-2\beta)(\widetilde{m}_i^2 + \widetilde{m}_j^2)) t},
\end{eqnarray}
\end{widetext}
for the QMUPL and the mass-proportional CSL models, respectively, where we have taken into account that the convolution $(g*g)(0) = (\sqrt{4\pi}r_C)^{-d}$. Finally, collecting all the integrals, we find the transition probabilities for the lifetime eigenstates and the flavor states,
\begin{widetext}
\begin{eqnarray}
    P_{M_i \rightarrow M_j}^{QMUPL}(t) &=& \Bigl( 1 - \lambda_Q \alpha (1-2\beta)\widetilde{m}_i^2 t\Bigr)^{-\frac{d}{2}} \delta_{ij}, \\
    P_{M^0 \rightarrow M^0/\bar{M}^0}^{QMUPL}(t) &=& \frac{1}{4}\Biggl\{ \sum_i \Bigl( 1 - \lambda_Q \alpha (1-2\beta)\widetilde{m}_i^2 t\Bigr)^{-\frac{d}{2}} \pm \frac{2\cos[t \Delta m]}{\Bigl( 1 - \frac{\lambda_Q \alpha}{2} \Bigl( (1-2\beta) \sum_i \widetilde{m}_i^2 - (\Delta \widetilde{m})^2 \bigr) t\Bigr)^{\frac{d}{2}}} \Biggr\}, \\
    P_{M_i \rightarrow M_j}^{CSL}(t) &=& e^{-\frac{\gamma}{(\sqrt{4\pi} r_C)^d} (2\beta - 1) \widetilde{m}_i^2 t} \delta_{ij}, \\
    P_{M^0 \rightarrow M^0/\bar{M}^0}^{CSL}(t) &=& \frac{1}{4}\Bigl\{ \sum_i e^{-\frac{\gamma}{(\sqrt{4\pi}r_C)^d}(2\beta - 1) \widetilde{m}_i^2 t} \pm 2 e^{-\frac{1}{2}\frac{\gamma}{(\sqrt{4\pi}r_C)^d} ((2\beta - 1) \sum_i \widetilde{m}_i^2 + (\Delta \widetilde{m})^2 ) t} \cos[t \Delta m] \Bigr\}.
\end{eqnarray}
\end{widetext}

\bibliographystyle{apsrev4-1} 
\bibliography{PRA_Published}

\end{document}